\documentclass{tMOP2e}

\usepackage{subfigure}
\usepackage{enumitem}
\usepackage[usenames,dvipsnames]{xcolor}
\usepackage[utf8]{inputenc}
\setlist[enumerate]{leftmargin=*,align=left,label=\thesection.\arabic*.}

\usepackage[pagebackref]{hyperref}	
\usepackage{units}
\hypersetup{colorlinks=true, linkcolor=blue, citecolor=blue, urlcolor=blue}

\theoremstyle{plain}

\theoremstyle{definition}

\theoremstyle{remark}

\newcommand{\ket}[1]{| #1 \rangle}

\begin{document}



\title{Quantum memories: emerging applications and recent advances}

\author{Khabat Heshami$^{\rm a}$$^{\ast}$, Duncan G. England$^{\rm a}$, Peter C. Humphreys$^{\rm b}$, Philip J. Bustard$^{\rm a}$, Victor M. Acosta$^{\rm c}$, Joshua Nunn$^{\rm b}$, Benjamin J. Sussman$^{\rm a,d}$ \thanks{$^\ast$Corresponding author. Email: khabat.heshami@nrc-cnrc.gc.ca
\vspace{6pt}}\\\vspace{6pt}  $^{a}${\em{National Research Council of Canada, 100 Sussex Drive, Ottawa, Ontario, K1A 0R6, Canada}}\\
  $^{b}${\em{Clarendon Laboratory, University of Oxford, Parks Road, Oxford OX1 3PU, UK}}\\
 $^{c}${\em{Dept. of Physics and Astronomy, University of New Mexico, Center for High Technology Materials, 1313 Goddard SE, Albuquerque, NM 87106, USA}}\\
$^{d}${\em{Department of Physics, University of Ottawa, Ottawa, Ontario, K1N 6N5, Canada}}
\received{\today} }

\maketitle

\begin{abstract}
Quantum light-matter interfaces are at the heart of photonic quantum technologies. Quantum memories for photons, where non-classical states of photons are mapped onto stationary matter states and preserved for subsequent retrieval, are technical realizations enabled by exquisite control over interactions between light and matter. The ability of quantum memories to synchronize probabilistic events makes them a key component in quantum repeaters and quantum computation based on linear optics. This critical feature has motivated many groups to dedicate theoretical and experimental research to develop quantum memory devices. In recent years, exciting new applications, and more advanced developments of quantum memories, have proliferated. In this review, we outline some of the emerging applications of quantum memories in optical signal processing, quantum computation, and nonlinear optics. We review recent experimental and theoretical developments, and their impacts on more advanced photonic quantum technologies based on quantum memories.
\end{abstract}



\section{Introduction}
It is broadly established that quantum mechanics has attributes not present in classical physics.  These unusual features --- like entanglement ---  can be used as a resource to construct technologies not historically understood to be possible.   Quantum sensors, quantum computers, and quantum cryptography all have specific enhancements over their classical counterparts \cite{giovannetti2011advances,ladd2010quantum,RevModPhys.74.145}.  The construction of these new quantum-enabled technologies, however, remains extremely challenging.  Out of the efforts to build quantum technologies has emerged the understanding that various quantum components will be essential, or very beneficial.  In the realm of photonics-based quantum technologies key quantum components include:  quantum memories, photon sources, frequency converters, quantum random number generators, and single-photon detectors.  The focus of this review is the quantum memory, a device that can store a single photon and recreate the quantum state~\cite{lvovsky2009optical,Simon2010,Tittel2010,Bussieres2013}. 

Quantum memories are under development by many groups around the world. Approaches to quantum memory encompass the full gamut of our understanding of electromagnetic interactions, and as such these research programmes represent the most advanced techniques for the quantum control of optical signals. For example resonant (first order) and Raman (second order) interactions in warm, cold, trapped, or Bose-condensed atoms, in amorphous and crystalline solids, molecular gases, structured media and metamaterials are actively pursued, with engineered couplings to electronic, magnetic, vibrational and hybrid degrees of freedom~\cite{Simon2010,Tittel2010,Bussieres2013}.

Photonics is a unique platform for quantum technologies because it can support broadband signals over long distances in ambient conditions, enabling, for example,  the first commercial quantum cryptographic devices. However, to extend the range of quantum cryptography systems, quantum memories are needed for repeaters~\cite{RevModPhys.83.33}.  More generally for quantum technologies, mechanisms are needed to generate and guide photons, to mediate non-linearities, and to detect photons. Quantum memories would enable the development of large photonic quantum processing systems, by providing the capability to coherently manipulate, buffer, and re-time photonic signals.

While the primary focus of quantum memory research has been the synchronization of entanglement swapping in quantum repeater protocols for long-distance quantum communication~\cite{RevModPhys.83.33}, the ability to interconvert material and optical quanta, to prepare non-classical states and read them out optically, and to generate and distribute long-lived entanglement, has led to a range of other applications~\cite{Bussieres2013}, so that \emph{quantum memories} is now a catch-all term for a broad class of research, linked by the common theme of coherent interfacing between light and matter~\cite{hammerer2010quantum,Nature.453.1023}.

As the field has developed, there have been outstanding review articles on quantum memories~\cite{lvovsky2009optical,Simon2010,Tittel2010,Bussieres2013}. We will endeavour to be complementary to these reviews. We will report on the most recent advances of the quantum memory community, and in particular we will discuss how the incorporation of spectral/temporal signal processing techniques from classical opto-electronics is leading to the development of new applications --- beyond quantum cryptographic repeaters ---  for quantum technologies based on quantum memory interactions.

The structure of the paper is as follows. In Sec.~\ref{sec:Implementations}, we discuss quantum memory protocols by summarizing existing schemes into approaches based on optical control of the light-matter interaction or engineered absorption. We then outline experimental developments based on several solid-state and atomic physical systems. As experimental efforts are moving past proof-of-principle demonstrations, it is crucial to underline challenges towards practical demonstrations of photonic quantum memories. We dedicate a subsection to discuss some of these implementation challenges~\ref{sec:Implementation_Challenges}. In Sec.~\ref{sec:application}, we elaborate on emerging applications of quantum memories that are complementary to those discussed in Ref.~\cite{Bussieres2013}. This includes applications of quantum memories in optical signal processing~\ref{sec:Processing}, optical quantum computation~\ref{sec:Computing} and non-linear interactions~\ref{sec:nonlinear}.

\section{Implementations}\label{sec:Implementations}
In the most general terms, the quantum memory process is the controlled absorption, and re-emission, of a photonic qubit; however, this generic definition hides a surfeit of experimental complexity. The practical implementation of a quantum memory requires the selection of a suitable medium in which to store the photon, and a protocol by which the user can manipulate the absorption and re-emission characteristics of the medium. Several protocols have been developed; each is briefly introduced in sub-section~\ref{sec:Protocols}. Quantum memory has been achieved in a range of physical systems which can be grouped into the following five platforms: rare-earth-ion doped solids, diamond color centers, crystalline solids, alkali metal vapors, and molecules. In sub-section~\ref{sec:Physical_Systems}, we discuss all of these physical systems and describe recent advances using each platform. 

When summarizing the various platforms and protocols, it is interesting to note that most protocols have been implemented on at least two different platforms, and most platforms can support multiple protocols. Despite this flexibility, no combination of protocol and physical system has yet provided all the desirable properties in a single package. Hence, the choice of system and protocol for a quantum memory will depend strongly on the application. For example, for long-distance communication, the memory storage time is of paramount importance, whereas for local quantum processing applications, other properties such as the time-bandwidth product, or the multi-mode capacity may be of more interest. In sub-sections~\ref{sec:Protocols} and \ref{sec:Physical_Systems} we aim to highlight the potential advantages of each system, and to highlight potential applications. 

In sub-section~\ref{sec:Implementation_Challenges} we conclude the overview of the state-of-the-art in quantum memory research by identifying three of the outstanding challenges in quantum memory research, and identify the on-going work toward solving these issues. 

\subsection{Protocols}\label{sec:Protocols}
Historically, quantum memories developed along two distinct paths: optically-controlled memories, and engineered absorption. In an optically-controlled memory, intense \textsl{control} pulses are used to mediate absorption and retrieval. Memories based on engineered absorption rely upon a large inhomogeneous broadening of an optical absorption line; this absorption can be structured, either statically or dynamically, to produce the desired absorption and re-emission characteristics. More recently, hybrid quantum memory schemes have been developed which combine engineered absorption and optical control to extract the beneficial features of both approaches. Below we summarize the various quantum memory protocols used in the literature:

\noindent {\em Optically-controlled memories ---} These memories are based upon a $\Lambda$-level system with a ground state $\ket{g}$, an excited state $\ket{e}$ and a metastable \textsl{storage} state $\ket{s}$ ({\em e.g.} see Fig.~\ref{fig:Noise}b). Dipole transitions are allowed from $\ket{g}$ and $\ket{s}$ to $\ket{e}$ but typically the transition from $\ket{g}$ to $\ket{s}$ is forbidden leading to a long radiative lifetime of the storage state. Absorption, and retrieval, of the weak \textsl{signal} pulse are each mediated by a strong \textsl{control} pulse. The signal and control pulses are in two-photon resonance with the storage state but each field can be detuned by an amount $\Delta$ from $\ket{e}$. In the regime where $\Delta$ is small compared to the linewidth $\Gamma$ of $\ket{e}$, the application of the control field results in a sharp dip in the absorption profile for the signal field, known as electromagnetically-induced transparency (EIT). EIT can be used to slow the group velocity of light~\cite{Harris1992}, and ultimately to store the light in the medium by dynamic adjustment of the control pulse intensity~\cite{PhysRevLett.84.5094}. When $\Delta\gg\Gamma$, despite the large detuning from optical resonance, absorption and retrieval can be achieved via off-resonant Raman scattering~\cite{Kozhekin2000,Nunn2007}. A unified theoretical discussion of these techniques can be found in~\cite{PhysRevA.76.033805}.

The re-emission from an optically-controlled memory only occurs when a control pulse is applied to the memory; thus the photon can be retrieved at any time, up to the coherence time of the memory. In this sense, an optically-controlled memory is truly \textsl{on-demand} which is a significant advantage, and sometimes a necessity, for many applications. The primary disadvantage of optically-controlled memories is that the intense control fields can introduce spurious noise photons which degrade the fidelity of the output. In addition, the control field used can restrict the storage of the signal field to a single optical mode such that multi-mode inputs can only be stored by spatial multiplexing~\cite{England2012}. However, it should be noted that, in some situations, this single mode selectivity may be advantageous, as will be discussed in section~\ref{sec:Computing}.   


\noindent {\em Engineered absorption ---} Motivated by the available inhomogeneous broadening in rare-earth ion doped crystals~\cite{Tittel2010}, several protocols have been designed based on controlling, or engineering, inhomogeneous broadening. This resulted in two of the most promising quantum memory protocols. In gradient echo memory (GEM)~\cite{PhysRevLett.96.043602,hetet2008electro}, an inhomogeneous absorption line is prepared, and then broadened by applying an electric or magnetic field gradient along the propagation direction. This is also known as longitudinal controlled-reversible-inhomogeneous-broadening (CRIB)~\cite{Kraus2006} that shares several similarities with \cite{PhysRevLett.87.173601,moiseev2003quantum}. After absorbing a signal photon in such a broadened ensemble, reversing the external field allows the ensemble to rephase at a certain point in time that determines the recall time. The atomic frequency comb (AFC) quantum memory protocol~\cite{PhysRevA.79.052329} relies on preparation of a periodic absorption feature, or comb, with equal peak spacing by optical pumping in an inhomogeneously-broadened ensemble. After absorption of the signal pulse, which spans multiple comb spacings, the inverse of this peak spacing pre-determines the rephasing time, and thus re-emission from the memory. This limitation of a fixed storage time is eliminated with optical control by transferring the stored excitation temporarily to an additional level \cite{PhysRevA.79.052329}.

Because these memories do not require a control field, they do not suffer from the same noise problems that afflict optically-controlled memories. Furthermore, because of the broadened absorption profile, they are intrinsically multi-mode, and can be used to store multiple temporal~\cite{Usmani2010,Bonarota2011} and spectral~\cite{PhysRevLett.113.053603} modes. However, readout from these memories cannot be considered truly on-demand: in CRIB, one must wait for the atomic polarization to rephase after the broadening has been reversed and, in AFC, the rephasing time is pre-determined by the comb spacing.  

\noindent{\em Hybrid Schemes ---} Motivated by the need for on-demand readout in engineered absorption quantum memories, research efforts have focused on the use of a `shelving state'. In these schemes, the signal pulse is absorbed in the structured, or broadened, medium as normal, an optical $\pi$-pulse then maps the excited state population into a meta-stable shelving state effectively pausing the procession of the atomic spins. After a controllable delay, a second $\pi$-pulse is applied and the population returns to the excited state, whereby the spin procession will continue as before, leading to re-emission of the light (see subsection \ref{spin_storage}). 

Optical $\pi$-pulses can be applied to an unstructured inhomogeneously broadened ensemble to rephase the collective atomic coherence generated by an absorbed single photon pulse. To avoid noise due to the inversed atomic population, the first echo can be silenced (suppressed) by spatial phase mismatch generated by the first optical $\pi$-pulse. A second $\pi$-pulse can bring the population back to the ground state and allow the ensemble to reach a second rephasing time with a proper phase matching to revive the silenced echo~\cite{damon2011revival}.

The optical preparation stage for an AFC can be used to achieve a spatio-spectral atomic comb, where in addition to temporal delay diffraction from the spatial grating allows to control readout direction of the signal \cite{PhysRevA.87.042338}. A combination of engineered inhomogeneous broadening and optical control has been used in \cite{kutluer2015spectral} to achieve storage in spin states of rare-earth-ion-doped ensembles. In this scheme, the authors used an optically-prepared spectral hole in an inhomogeneously broadened ensemble to reduce the group velocity of an optical pulse, such that the pulse is spatially-compressed into their crystal. Then a short Raman pulse is used on transfer this optical excitation to a spin state for storage. Spin storage has been achieved in several recent experiments that are discussed in subsection~\ref{spin_storage}.

\noindent{\em Recent protocols ---} Over the past few years several theoretical schemes have been proposed that led to a better understanding of inherent similarities between different quantum memory protocols, and minimal requirements for a storage and retrieval process. Here we summarize some of these protocols and their connection to already existing schemes.

Control over the transition dipole moment in a two-level system has been shown theoretically to allow storage and retrieval of photons equivalent to an off-resonant Raman quantum memory~\cite{heshami2012controllable}. This was inspired by the possibility of modulating the transition dipole moment in Tm:YAG by applying an external magnetic field~\cite{PhysRevB.71.174409}. In another proposal, sweeping the transition frequency of a two-level system enabled storage and retrieval of photons, such that it resembles some of the characteristics of EIT and GEM~\cite{NJP.15.085029}. Another approach in an ensemble of two-level systems involving two species with a controlled homogeneous splitting led to similarities to EIT~\cite{hetet2013quantum}. In addition, a space-dependent
frequency shift of a propagating signal, as can be induced by a time-dependent refractive index, has also been shown to be equivalent to the gradient echo memory~\cite{PhysRevA.86.013833}. 

 Several schemes have been proposed to use different degrees of freedom of light to induce longitudinal CRIB. For example, phase-matching control~\cite{PhysRevA.87.013811} or spatial chirp of a control field~\cite{PhysRevA.83.053849,PhysRevA.90.052322} have been shown to reproduce the results of the gradient echo memory that was originally associated to controlled inhomogeneous broadening. These are in contrast to the temporal control induced by the shape of the control field in off-resonant Raman and EIT storage schemes.


\subsection{Physical systems and their recent developments}\label{sec:Physical_Systems}
\subsubsection{Rare-earth-ion doped solids}
Rare-earth-ion doped solids are appealing candidates for implementing elements for optical quantum information processing~\cite{thiel2011rare}. They possess narrow homogeneous linewidths, provide optical access to electronic and nuclear spin states, and demonstrate significant optical inhomogeneous broadening that can be tailored by optical pumping techniques. In addition, these systems allow integrable \cite{nature.469.512,PhysRevLett.115.013601} and micron/nano scale \cite{Faraon2015} implementations. Rare-earth ion doped crystals have initially been used for implementing storage of time-bin qubits using the AFC protocol~\cite{PhysRevA.79.052329}. Given that the dipole moments are aligned along a specific direction in the solid, two quantum memories can be used for polarization qubit storage. This has been demonstrated using different configurations in \cite{PhysRevLett.108.190503, PhysRevLett.108.190504, PhysRevLett.108.190505}; and more recently for temporally-multiplexed polarization qubits in \cite{laplane2015multiplexed}. In addition, Jin \emph{et al.}~\cite{ PhysRevLett.115.140501} used an ensemble of erbium atoms in an optical fiber for direct storage and retrieval of heralded photonic polarization qubits at telecommunication wavelength. In a more recent experiment, quantum storage of orbital-angular-momentum (OAM) entanglement was achieved~\cite{ PhysRevLett.115.070502}. The authors employed the AFC protocol in Nd$^{3+}$:YVO$_4$ and used the transverse spatial degree of freedom to store OAM entangled states. This demonstrates the spatial multimode capacity of rare-earth-ion-based light-matter interfaces beyond their already existing capability to process temporally- \cite{Usmani2010,1367-2630-15-4-045012,Bonarota2011,laplane2015multiplexed,Tang2015} and spectrally-multiplexed \cite{PhysRevLett.113.053603,NJP.16.065019} states. Below, we outline some of the most recent developments in rare-earth-ion-based quantum memories.

\paragraph{Spin storage and long coherence time in solids}\label{spin_storage}
One of the main advantages of rare-earth-ion doped crystals is their excellent coherence properties at cryogenic temperatures. This is particularly important for application of quantum memories in long-distance quantum communications. In order to enable entanglement distribution beyond distances achievable by direct transmission in fibers ($>$1000 km), storage times in excess of milliseconds will be required.

In one of the early experiments based on rare-earth-ion doped crystals \cite{ PhysRevLett.95.063601} storage times of greater than one second were demonstrated for classical signals in Pr$^{3+}$:Y$_2$SiO$_5$ using the \textsl{bang-bang} dynamical decoupling technique. Extending this approach to the quantum level was suggested to require a precision in control $\pi$-pulses of order $1/N$, where $N$ is the total number of atoms~\cite{PhysRevA.70.032320}.

However, thanks to collective directional emission from the atomic ensemble, it has been shown that acceptable signal to noise ratios are possible for single photon storage where spin-echo technique is implemented using RF $\pi$-pulses to extend the spin coherence time beyond the limit imposed by the spin inhomogeneous broadening~ \cite{PhysRevA.83.032315}. Several other dynamical decoupling sequences are characterized from this perspective in a recent study~\cite{cruzeiro2015noise}. Similar spin rephasing experiments have been performed at the quantum level in cold atomic ensembles \cite{PhysRevLett.115.160501,PhysRevLett.115.133002}. The applicability of dynamical decoupling techniques in solid-state spin wave storage was experimentally demonstrated in Pr$^{3+}$:La$_2$(WO$_4$)$_3$ \cite{PhysRevLett.111.020503}. In a more recent experiment, spin-echo manipulation with a mean excitation number of one was performed in Eu$^{3+}$:Y$_2$SiO$_5$ \cite{PhysRevLett.114.230502}. After a storage time of about 1 ms, the authors managed to optically retrieve the spin excitation with high signal-to-noise ratio. Contemporaneously, authors in \cite{PhysRevLett.114.230501} achieved spin-wave storage of time-bin qubits with low noise in Pr$^{3+}$:Y$_2$SiO$_5$. Combination of these approaches will be required for entanglement distribution over long distances with quantum repeaters. 

Long coherence times have been shown in Pr$^{3+}$:Y$_{2}$SiO$_{5} $ \cite{ PhysRevLett.95.030506}; however, it is important to demonstrate a long coherence time combined with optical storage and retrieval. In a more sophisticated experiment \cite{ Heinze2013}, a self-learning evolutionary optical pumping technique and an improved dynamical decoupling method were employed to extend EIT storage time in the same crystal up to one minute. 

The highly influential work of Zhong {\it et al.} \cite{ Zhong2015} led to a recent breakthrough in extending the coherence time of ground-state hyperfine transitions in $^{151}$Eu$^{3+}$:Y$_2$SiO$_5$ to 370$\pm$60 minutes at 2~K. Such a long coherence time, combined with storage of photonic entanglement, allows one to re-think long-distance entanglement distribution approaches.

\subparagraph{Microwave storage and optical to microwave conversion}
Electronic and nuclear spin states in rare-earth-ion doped crystals enable direct storage and retrieval of microwave photons – a feature that can be used to compensate for the lack of long-term storage in superconducting qubits. Spin ensembles in Nd$^{3+}$:YSO~\cite{ PhysRevLett.114.170503} and Er$^{3+}$:YSO~\cite{PhysRevB.92.014421} have been used to store microwave photons. The nuclear spin in Nd$^{3+}$:YSO offers a coherence time of 9~ms at $\approx$ 5\,K that is approximately 3 orders of magnitude longer than electronic spin coherence time of Er ions in YSO (below 1\,K), where magnetic dipole-dipole interaction between the large magnetic moments of Er$^{3+}$ ions is believed to be the primary source of dephasing \cite{bottger2003material}. The optical transition in Er$^{3+}$:YSO near the telecommunication C band at 1.54~$\mu$m is a unique feature that promises coherent interconversion between microwave and telecom photons.  In general, optical access to spin states in rare-earth-ion doped crystals precipitates hybrid quantum devices for coherent coupling between optically-excited spins and superconducting resonators. In \cite{PhysRevLett.110.157001}, strong coupling between magnetically anisotropic Er$^{3+}$:Y$_2$SiO$_5$ crystals and superconducting resonators was demonstrated. Microwave storage has also been shown in single-crystal YIG spheres inside a three dimensional microwave cavity. This is achieved by controlling local magnetic field of each YIG sphere to engineer the coupling between magnon modes and microwave photons~\cite{zhang2015magnon}. These demonstrations are step towards optical to microwave photon interfaces~\cite{PhysRevLett.113.063603,PhysRevLett.113.203601} in hybrid architectures (see also \ref{NV_centers}).

\paragraph{Cavity-enhanced implementations}
In spite of a few demonstrations of efficient solid-state~\cite{nature.465.1052} and atomic~\cite{ncomms.2.174} memories, achieving high efficiency operation has always been challenging. This may become more difficult if one aims to combine high efficiency with multimode functionality and long storage time. Employing optical cavities is a known approach to implement more efficient light-matter interfaces. This has been shown for quantum memories based on $\Lambda$-level systems \cite{PhysRevA.76.033804}. In particular, cavity enhancement for rare-earth-ion doped crystals has been shown theoretically for the AFC storage protocol \cite{PhysRevA.82.022310, PhysRevA.82.022311}. This led to experimental demonstration of an efficient quantum memory in optically thin atomic ensembles \cite{PhysRevLett.110.133604, jobez2014cavity}. Recently, Zhong {\it et al.} \cite{Faraon2015} demonstrated a mesoscopic ensemble of rare-earth Nd$^{3+}$ ions in YSO coupled to a high quality-factor nano-cavity. This will provide a platform for rare-earth-ion-based nanophotonics with multimode capacity, potentially high efficiency, access to hyperfine states for long coherence times, and the possibility of on-chip multiplexed implementations for miniaturized architectures.

\subsubsection{Nitrogen-vacancy centers in diamond}\label{NV_centers}
Nitrogen-vacancy (NV) centers in diamond motivated a great deal of research during the past decade due to their excellent properties for optical micro/nano-photonic devices. The possibility for optical preparation and readout of electronic and nuclear spin states~\cite{dutt2007quantum, gaebel2006room}, and long electronic and nuclear spin coherence times~\cite{Maurer2012, bar2013solid} have led to exciting experiments~\cite{togan2010quantum, bernien2013heralded, pfaff2014unconditional, hensen2015experimental}. Raman excitation of spin coherences and coherent population trapping have been demonstrated in an ensemble of NV centers \cite{hemmer2001raman, santori2006coherent}. In a more recent experiment~\cite{PhysRevLett.110.213605}, EIT was implemented in a multi-pass diamond chip to achieve all-optical electromagnetic field sensing. Despite these closely-related experiments, optical storage remains to be achieved in practice. The available level structure of NV centers (both negatively charged and neutral NV centers) makes optical storage possible in NV centers in diamond~\cite{heshami2014raman, poem2015broadband}. 

\paragraph{Microwave storage and optical to microwave conversion}  Electronic spin states of NV centers enable microwave photon storage~\cite{kubo2012storage, PhysRevX.4.021049, PhysRevA.92.020301, grezes2015storage}. Coupling between NV spin ensembles and superconducting qubits~\cite{kubo2011hybrid,marcos2010coupling,kubo2010strong,zhu2011coherent} promises microwave storage for superconducting qubits. Optical storage combined with coupling of the electronic spin states to superconducting qubits allow NV centers in diamond to function in hybrid quantum systems for coherent optical to microwave interconversion~\cite{PhysRevA.91.033834}. 

\subsubsection{Raman scattering in solids}

Diamond has a high Raman gain on its optical phonon mode at 1,332\,cm$^{-1}$ (40\,THz)~\cite{Solin1970}, and a broad transimission window in the visible and near-infrared, making it suitable for use as an extremely broadband optical memory. The large Raman shift, combined with inhibited four-wave mixing due to material dispersion, means that it has quantum-level noise properties, even at room temperature~\cite{PhysRevLett.111.243601}. After the Raman storage interaction, the optical phonon decays into a pair or acoustic phonons via the Klemens channel~\cite{Klemens1966,Liu2000}; the lifetime of this decay is 3.5\,ps~\cite{Waldermann2008,Lee2010} which makes the diamond memory unsuitable for communication protocols and many processing applications. Nevertheless, the diamond memory has allowed several revealing studies of light-matter interactions at the quantum level: optical phonons in diamond have been used to demonstrate an emissive quantum memory~\cite{Lee2012}, macroscopic entanglement~\cite{Lee2011a}, storage of heralded single photons~\cite{PhysRevLett.114.053602}, and frequency/bandwidth manipulation of single photons~\cite{Fisher2015}.

In a related recent work, a modified Raman storage interaction was used to demonstrate optically-controlled, continuously-tunable slow light with femtosecond-duration pulses in a potassium titanyl phosphate waveguide~\cite{1508.01729}. Signal pulses were slowed by the dispersion created when a narrowband control pulse was applied near two-photon Raman resonance with a Raman absorption doublet.

\subsubsection{Alkali vapours} 

A large resonant optical depth is a prerequisite for the construction of an efficient quantum optical memory~\cite{PhysRevA.76.033805}. Alkali metal vapour isotopes have large optical depths at near-infrared wavelengths because of their relatively-narrow spectral lines and high number densities at `warm\rq{} temperatures of 50-100$^\circ$C. These atoms also have low-lying spin states which do not fluoresce, and have long coherence times; ensemble coherence lifetimes in warm alkali vapours are frequently limited not by intrinsic coherence times, but by atomic motion removing atoms from the optical interaction region, disturbing the spatial coherence of ensemble excitations, or causing collisional linewidth broadening. Much longer lifetimes can be achieved by cooling atoms to $\mu$K-temperatures in a magneto-optical trap (MOT)~\cite{Phillips1998} or confining them in an optical dipole trap~\cite{Grimm2000}. Alkali vapour spin states therefore make excellent `shelf\rq{} or storage states for $\Lambda$-level protocols such as EIT~\cite{PhysRevLett.86.783}, $\Lambda$-GEM~\cite{JPhysB.45.124004}, and Raman-type schemes~\cite{Reim2010}, whether at warm or ultracold temperatures. Thanks to the attributes of high optical depth, long coherence times, and easily-accessible optical transitions in the near-infrared, alkali vapours have been used for some of the most important memory developments, ranging from early research~ \cite{Nature.423.731,Lukin2003}, to the most recent achievements, to which we now turn.

\subparagraph{Orbital angular momentum storage in alkali vapours}

There has been increasing interest in the use of spatially-structured photons for quantum information purposes, due to their high capacity for information transmission. Structured light carries OAM, which must be stored if a memory is to faithfully reproduce a stored, structured photon. Atomic vapour quantum memories are well-suited to the storage of such beams because the OAM of a photon can be mapped into the phase and amplitude of a distributed ensemble excitation. Diffusion is the chief limitation of this technique, because thermal atomic motions destroy the spatial coherence of the stored excitation. Early successes involved the storage of weak, coherent pulses with spatial structure, in warm~\cite{PhysRevLett.100.123903,PhysRevLett.100.223601,OptExp.20.12350,PhysRevA.86.023801,chrapkiewicz2012generation}, and ultracold~\cite{PhysRevA.87.013845}, atomic ensembles. This was followed by the storage of single-photon-level Laguerre-Gaussian azimuthal modes, with $l=\pm1$ OAM units, in a cesium MOT using EIT~\cite{OptLett.38.712}. Using the same system, qubit storage and retrieval was demonstrated for coherent states with mean photon number $\bar{n}=0.6$, a raw average fidelity of $92.5\pm2\%$, and efficiency $\eta=15\pm2\%$~\cite{NatPhoton.8.234}. More recently, the same group was able to store and retrieve vector beams -- characterized by variation of the polarization in the transverse beam-plane -- at the single-photon level, in a dual-rail cesium MOT~\cite{NatCommun.6.7706}. The memory conserved the rotational invariance of the vector beams, making it a possible candidate for use with qubits encoded for misalignment-immune quantum communications~\cite{NatCommun.6.7706}.

The first storage of structured, true single photons was achieved using EIT in a rubidium MOT~\cite{NatCommun.4.2527}. Heralded single photons produced by spontaneous four-wave mixing in one MOT were prepared with one unit of OAM using a spiral phase plate, stored in a second MOT, and retrieved; the recalled photons violated a Cauchy-Schwarz inequality, verifying their preserved non-classical characteristics. Using a dual-rail setup the authors were also able to demonstrate the coherence of the multimode memory by storing the OAM superposition state of a heralded single photon for 100~ns. More recently, the same group demonstrated storage and retrieval of OAM entanglement in two separate rubidium ensembles (MOT~A and MOT~B)~\cite{PhysRevLett.114.050502}. An anti-Stokes single photon (signal 1) was created by spontaneous Raman scattering in MOT~A and then stored in MOT~B by the Raman memory protocol, creating OAM-entanglement between the two atomic ensembles. The authors retrieved the excitations from MOT~A and MOT~B using the Raman memory protocol; the OAM entangled photons were used to violate the CHSH inequality~\cite{PhysRevLett.23.880} verifying that OAM entanglement was maintained throughout storage of the signal photon in MOT~B. Using a similar design, the same group subsequently demonstrated post-selected polarization entanglement storage between two MOTs~\cite{NatPhoton.9.332}. 

\subparagraph{Broadband storage in alkali vapours}

The demonstrations of true single-photon Raman storage in atomic vapours~\cite{NatCommun.4.2527,PhysRevLett.114.050502,NatPhoton.9.332} represent an important advance, because the Raman scheme can store exceptional bandwidths~\cite{PhysRevLett.111.083901,PhysRevLett.114.053602}. The highest bandwidth atomic vapour memory to date has been demonstrated in cesium~\cite{Reim2010a}; however, efforts to demonstrate GHz-bandwidth, true single-photon storage in cesium have been limited by deleterious four-wave mixing (FWM) which introduces noise photons~\cite{NJP.17.043006}. FWM noise may be suppressed by placing the cesium vapour in a hollow-core fibre~\cite{NJP.15.055013,NatPhoton.8.287}, or in a low finesse cavity~\cite{arxiv1510.04625v1}. In the case of a cavity, the efficiency is enhanced by overlapping the signal frequency with a cavity resonance feature; the noise is suppressed by overlapping the noise photon frequency with an anti-resonance feature~\cite{arxiv1510.04625v1}.

\subparagraph{Long lifetimes in alkali vapours}

Atomic motions are one of the main causes of loss in vapour memories, either destroying ensemble coherence, or simply through loss of atoms from the interaction region. Cooling atoms in a MOT significantly reduces atomic speeds, enabling $\mu$s memory lifetimes, which are still insufficient for long-distance repeater applications, or when a large time-bandwidth product is required. When applied to cold atoms, optical lattices have been shown to limit atomic motions, enabling ms-duration storage times in rubidium, when combined with a ``magic\rq{}\rq{} magnetic field to compensate for differential Stark shifts~\cite{nphys.5.100}. Recently, it was shown that a microwave dynamic decoupling sequence can be used to further extend the coherence time of the rubidium clock transition to 16~s for atoms in an optical lattice~\cite{PhysRevA.87.031801}; this is a significant improvement compared with untrapped atomic vapours.

\subparagraph{Alkali vapours in cavities and fibers}

One of the key challenges with alkali vapour memories is to efficiently couple photons to and from preferred spatial modes; for example, the ability to integrate memories with optical fibre modes is critical for many quantum communication schemes. One possibility is to place atoms in a hollow-core fibre~\cite{NJP.15.055013,NatPhoton.8.287}; however, adsorption of atoms to the fibre wall reduces the optical depth, and atomic collisions limit the coherence time. An alternative route is to place a solid-core nanofibre with sub-wavelength diameter in an ultracold atomic cloud, such that the evanescent wave of light guided by the fibre interacts with the atoms~\cite{PhysRevA.66.063808}. These two possible routes to fiber integration of atomic vapour memories are discussed in detail in section \ref{sec:Integration}. Optical cavities have also been used to demonstrate efficient coupling between atomic vapour memories and the cavity mode, with the cavity increasing the coupling strength~\cite{PhysRevLett.98.183601,NatPhys.8.517}. Recently, one group demonstrated efficient retrieval of an entangled spin-wave qubit placed in a ring-cavity~\cite{PhysRevLett.114.210501}. The spin-wave qubit was prepared by Raman scattering on the hyperfine levels of Rb$^{87}$, with Stokes photons emitted into the cavity mode; after a delay, a read pulse induced emission of an anti-Stokes signal photon with net retrieval efficiencies of up to 17\%. Joint polarization measurements on the Stokes and signal photons demonstrated entanglement between the photons which violated a Bell-type inequality~\cite{PhysRevLett.114.210501}. Technical improvements may allow the net retrieval efficiency to approach the demonstrated intrinsic retrieval efficiency of 76(4)\%, making this potentially an attractive source for quantum communication applications.

\subsubsection{Molecules for storage and processing}

Most memory research has used atoms, or atomic ions, as the storage medium. As an alternative, molecules offer a rich level structure, which may be better-suited to specific applications. For example, hydrogen has its first optical resonance at ultraviolet wavelengths, such that broadband optical pulses can propagate through the gas with very low absorption and little dispersion. Recently, a room temperature coherent absorptive Raman memory was demonstrated in hydrogen using the first vibrational level as the storage state; weak, coherent 100-fs duration pulses at \unit[600]{nm} were stored for durations approaching \unit[1]{ns}~\cite{PhysRevLett.111.083901}. In a subsequent work, the authors demonstrated an emissive quantum memory on the rotational levels of hydrogen, using the DLCZ memory protocol. Nonclassical correlations were measured between a spontaneously emitted Stokes photon and a retrieved anti-Stokes photon, each of $\approx\unit[150]{fs}$ duration, and separated by up to \unit[90]{ps}~\cite{OptLett.40.922}. These results demonstrate the promise of molecules for ultrafast local quantum processing, where long storage times are not required.

In another experiment using molecules, a spin echo memory was demonstrated using tetracyanoethylene anion radicals (TCNE$^{-}$) in a toluene solution~\cite{PhysRevA.90.042306}. An applied magnetic field was used to prepare an AFC-type spin-frequency comb on the \unit[9.612]{GHz} hyperfine absorption line of TCNE$^-$ placed in a microwave cavity. The authors were able to demonstrate coherent absorption and echo retrieval of microwaves.

\subsection{Implementation challenges}\label{sec:Implementation_Challenges}
\subsubsection{Single photon storage}

Many proof-of-principle optical memory demonstrations are made using bright laser pulses containing multiple photons ({\em e.g.} refs.~\cite{PhysRevLett.111.083901,Cho2010,Reim2010}). However, to leverage the full power of quantum communication and processing, quantum memories must faithfully store true single photons. Single photon storage presents two significant challenges above and beyond bright pulse storage. Firstly, at the single photon level, spurious noise photons can contaminate the memory output and degrade the fidelity of the retrieved single photon. Secondly, generating single photons whose wavelength and bandwidth are well-matched to quantum memories is technically difficult. These two challenges, and experimental progress toward overcoming them, are discussed below.

\paragraph{Noise}

\begin{figure} 
\center{\includegraphics[width=0.9\linewidth, viewport=0 490 600 650]{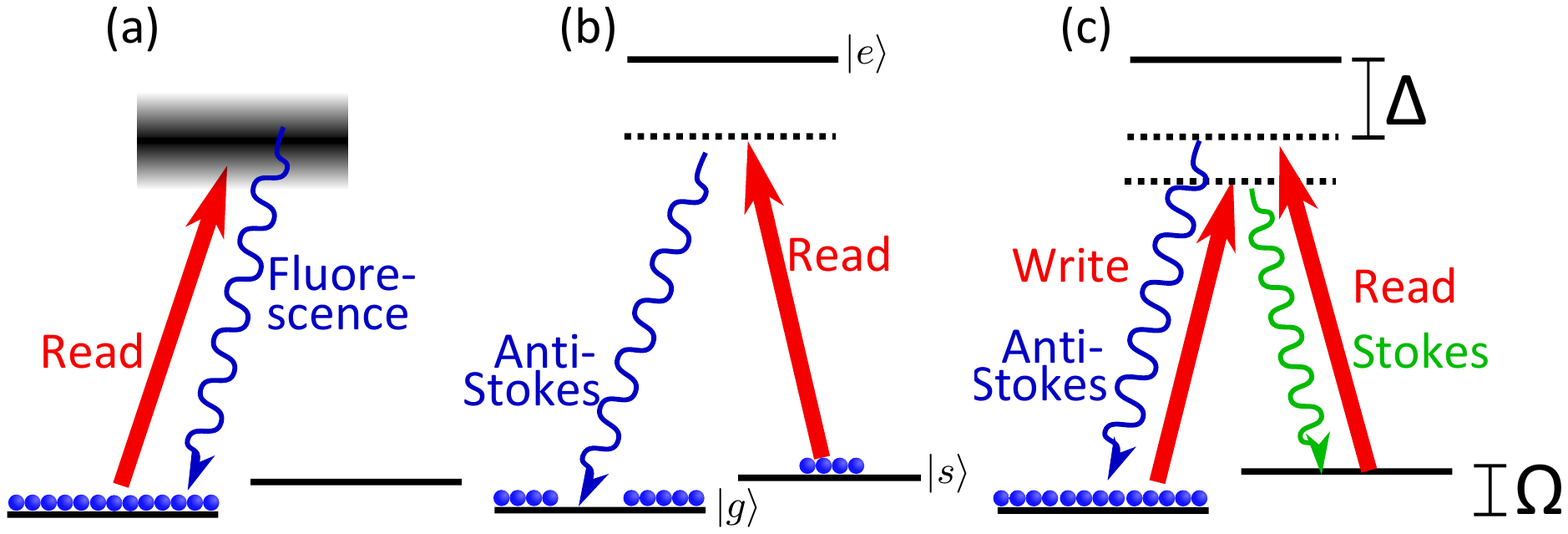}}
\caption{Prevalent noise processes in quantum memories. {\bf (a) Resonant fluorescence:} In near-resonant schemes the read pulse can excite population in the intermediate state $\ket{e}$. Doppler broadening of this line can result in fluorescence at the signal frequency.  {\bf (b) Spontaneous Raman scattering from thermal population:} Thermal excitation can populate the storage state, anti-Stokes Raman scattering from this thermal population is a source of noise. {\bf(c) spontaneous four-wave mixing:} The write pulse undergoes spontaneous Stokes scattering to generate a Stokes photon and a material excitation. The read pulse scatters from the excitation generating a photon at the anti-Stokes frequency and returning the population to the ground state. The Stokes photon can be eliminated by spectral filtering, but the anti-Stokes photon is a source of noise. }
\label{fig:Noise}
\end{figure}

Optically-controlled quantum memories such at EIT or Raman-coupled protocols require intense read and write pulses to mediate the storage and retrieval interactions. Read/write pulses typically contain between $10^6$~\cite{PhysRevLett.114.180503} and $10^{13}$~\cite{PhysRevLett.111.083901} photons and, in many cases, are extremely close in frequency to the single photon input\footnote{In rubidium quantum memories, the frequency difference between signal and read/write is just 6.8\,GHz~\cite{ncomms.2.174}}. Filtering out this strong field, while leaving the signal field unaffected, is a difficult but essential task, which can be achieved by a combination of spatial, spectral, and polarization filtering~\cite{Kupchak2015,dkabrowski2015magnetically}. However, in many cases, the intense read/write pulses can generate noise photons, by various linear and non-linear optical processes, which cannot be removed by filtering. The three dominant sources of noise are resonant fluorescence, thermal population of the storage state, and spontaneous four-wave mixing; these mechanisms are shown in figure~\ref{fig:Noise}, and are outlined below.

\subparagraph{Fluorescence noise}
If the read pulse is near-resonant with the $\ket{g}\rightarrow\ket{e}$ transition, then it can excite population into the intermediate state $\ket{e}$. Population in $\ket{e}$ will spontaneously decay back to the ground state $\ket{g}$, re-emitting a photon. In a Doppler-broadened atomic medium, the fluorescence photon can be shifted in frequency due to a change in atomic velocity, generating noise photons at the signal frequency (see figure~\ref{fig:Noise}(a)). Fluorescence noise is a particular problem in `warm' memories where the Doppler linewidth of $\ket{e}$ is typically $\gtrsim500$\,MHz; large detunings are therefore required to eliminate fluorescence. In ultracold or cryogenic memory substrates, the extremely narrow Doppler linewidths allow near-resonant operation while eliminating fluorescence. 

Fluorescence noise decays exponentially with half-life determined by the linewidth of $\ket{e}$ (typically $\sim10$\,MHz); therefore, if short pulses ($\lesssim 1$\,ns) are stored, the effect of fluorescence noise can be mitigated by time-gating~\cite{Reim2011PRL}. An alternative method to remove fluorescence noise is to detune signal and read/write pulses from $\ket{e}$ while maintaining two-photon resonance between $\ket{g}$ and $\ket{s}$. The memory coupling strength decreases as the detuning from $\ket{e}$ is increased; higher-power read/write pulses can be used to compensate for this, although the increased power may induce deleterious Raman noise effects (see below).

\subparagraph{Thermal population of the storage state}

In the generic $\Lambda$-level structure of figure~\ref{fig:Noise}(b), the metastable state $\ket{s}$ is $\hbar\Omega$ higher in energy than the ground state $\ket{g}$. The thermal ratio of populations at temperature $T$ is given by the Boltzmann factor $e^{-\hbar\Omega/k_BT}$. For a thermal sample, unless $\hbar\Omega\gg k_BT$, the read pulse will scatter from population in $\ket{s}$, creating anti-Stokes noise photons with the same frequency, polarization, and temporal structure as the signal photons.


To reduce thermal anti-Stokes noise, the storage state must be nearly empty before the write interaction. It is therefore desirable to have a small Boltzmann factor, implying either large $\Omega$ or low temperature. In atomic systems, $\ket{g}$ and $\ket{s}$ are often hyperfine states within the electronic ground state and thus typically $\Omega \lesssim 10$\,GHz. Therefore, at room temperatures, the Boltzmann factor approaches one such that $\ket{g}$ and $\ket{s}$ are equally populated. In this case, $\ket{s}$ must be emptied by optical pumping. High bandwidth Raman memories in diamond~\cite{PhysRevLett.111.243601,PhysRevLett.114.053602} and hydrogen~\cite{PhysRevLett.111.083901} have large $\Omega$: 40\,THz and 125\,THz respectively. Therefore, even at room temperature, these systems require no optical preparation for use in memories.

\subparagraph{Four-wave mixing (FWM)}
The FWM process is illustrated in figure~\ref{fig:Noise}(c); a photon from the write pulse\footnote{For the purpose of this illustration, we assume that the first photon in the FWM process originates from the write pulse. In reality, it could also come from the read pulse.} is spontaneously scattered to produce a Stokes photon and a material excitation. A photon from the read pulse then scatters from the material excitation to produce an anti-Stokes photon. The Stokes photons can be removed by spectral filtering, but the anti-Stokes photons are impossible to distinguish from the signal~\cite{Lauk2013}.

As can be seen in figure~\ref{fig:Noise}(c), if the memory is red-detuned from state $\ket{e}$ by $\Delta$ then the first step of the SFWM (Stokes scattering) is detuned by $\Delta + \Omega$. In the case when $\Omega \gtrsim \Delta$ the memory interaction experiences stronger Raman coupling than the SFWM, due to the smaller detuning, so FWM noise is suppressed. It is therefore desirable to have a large separation between ground states and to operate as close to resonance with the intermediate state as can be allowed by fluorescence noise. In the limit where $\Omega \ll \Delta$ --- for example in far-off-resonant Raman protocols~\cite{Reim2010,PhysRevLett.111.083901,PhysRevLett.114.053602} --- the Raman coupling strength for FWM becomes comparable to that of the memory and hence FWM can be a significant source of noise. In these systems other methods, such as polarization selection rules~\cite{Zhang2014}, phase-matching in a dispersive medium~\cite{PhysRevLett.111.243601}, or non-colinear geometry~\cite{dkabrowski2014hamiltonian}, can be used to minimize the detrimental effects of FWM.

\paragraph{Single Photon Generation}

Single photon generation is an active area of research spanning a range of different platforms~\cite{Eisaman2011}. Broadly, single photon sources can be split into two types: deterministic and heralded. Deterministic single photon sources are based on single emitters, such as semiconductor quantum dots~\cite{Kako2006}, single molecules~\cite{Lounis2000}, single atoms~\cite{McKeever2004}, single ions~\cite{Keller2004} or diamond color centers~\cite{Kurtsiefer2000}. The wavelength and bandwidth of these deterministic sources is defined by the transition wavelength and the natural linewidth of the emitter, therefore deterministic photon sources typically emit narrow-bandwidth photons at a single well-defined frequency. Heralded single photon sources generate correlated pairs of photons via a spontaneous processes; either spontaneous parametric downconversion (SPDC) in $\chi^{(2)}$ non-linear crystals~\cite{PhysRevLett.25.84} or spontaneous four-wave mixing (SFWM) in $\chi^{(3)}$ optical fibers~\cite{Goldschmidt2008,OptExpress.17.23589} or waveguides~\cite{Takesue2007,Spring2013a}. The photon pairs are then split on a beam splitter and the detection of a \textsl{herald} photon exiting one beam splitter port implies the existence of the \textsl{signal} photon exiting the other port. Because SPDC and SFWM do not rely upon optical dipole transitions, the wavelength of photon pair emission is broadly tuneable. The bandwidth generated by heralded sources is generally large because energy and momentum can be conserved over a broad bandwidth ($\sim$THz).

The majority of quantum memories operate close to an optical dipole transition and therefore can only store photons of a narrow bandwidth ($\lesssim 5$\,GHz) at a fixed wavelength defined by the dipole transition. Matching the bandwidth and the wavelength of a single photon source and a quantum memory is a key challenge. The bandwidths of single-emitter photon sources (~$\sim$MHz) are well suited to storage in quantum memories, but they typically have fixed frequencies that are difficult to tune into resonance with the quantum memory. The frequency of quantum dot single emitters can be tuned by the application of external electric and magnetic fields~\cite{Akopian2010}. This technique has been used to great effect, tuning a single GaAs quantum dot into resonance with rubidium vapor for slow-light~\cite{NatPhoton.5.230}. Similarly, local heating of a InAs/GaAs quantum dot was used to tune its emission into resonance with a Nd$^{3+}$ ions embedded in a YVO$_4$ crystal allowing storage of multiple single photons in an AFC quantum memory~\cite{Tang2015}. The emission linewidth of a quantum dot photon source can be artificially narrowed by constructing an optical cavity around the dot. By adjusting the cavity the emission frequency can be tuned, within the natural linewidth, in order to match the frequency to a chosen transition. This technique was illustrated by the direct coupling of an InAs quantum dot with a $^{174}$Yb$^+$ ion~, despite a 60-fold mismatch of the radiative lifetimes of the two systems~\cite{Meyer2015}. In reference \cite{siyushev2014molecular}, Siyushev et.al. explore the use of complex molecules as single emitters. They use a cryogenic sample of the organic laser dye dibenzanthanthrene (DBATT, C$_{30}$H$_{16}$) which has a resonance well-matched to the Sodium D-line transitions at 589\,nm and a transform-limited linewidth of 17\,MHz. The inhomogeneous broadening in the cryogenic sample is enormous ($>1$\,THz), ordinarily, this is a disadvantage but here it is an advantage because, out of a sample of several thousand DBATT molecules, a few can be found that are exactly in resonance with the atomic line. One of these molecules is then imaged using a microscope to act as a single emitter. This technique was used to crease single photons resonant with the Sodium D$_1$ and D$_2$ lines and, by using different molecules, the D-lines in both Potassium and Rubidium.

 In contrast to single emitters, heralded photon sources are highly tunable, making it simple to tune the photons into resonance with a quantum memory, however the large photon linewdith ($\sim$THz) is incompatible with the bandwidth of most quantum memories ($\lesssim 5$\,GHz). Four different approaches designed to bridge the bandwidth gap between sources and memories are detailed below.

\begin{enumerate}[label=(\alph*)]
\item {\bf Filter the photon source:} Tight spectral filtering can be applied to the photons. This has been used to match the bandwidth of SPDC photon sources with AFC memories in Nd$^{3+}$:Y$_2$SiO$_5$~\cite{nature.469.508,PhysRevLett.108.190503}, a Ti:Tm:LiNbO$_3$ waveguide~\cite{Saglamyurek2012,nature.469.512} and erbium-doped fiber~\cite{NatPhoton.9.83} as well as a Raman memory in cesium vapour~\cite{NJP.17.043006}.

\item {\bf Build the source in a cavity:} Optical cavities built around spontaneous parametric downconversion sources can reduce the emission linewidth and increase tunability~\cite{PhysRevLett.101.190501,Scholz2009,Wolfgramm2011}. This technique has been used to address an AFC memory in Pr$^{3+}$:Y$_2$SiO$_5$~\cite{Fekete2013,Rielander2014} and an EIT memory in ultracold rubidium~\cite{Zhang2011}. Tunable MHz-bandwidth single photon sources have recently developed using SPDC in whispering-gallery mode resonators~\cite{Schunk2015} and SFWM in micro-ring resonators~\cite{Ramelow2015}, though these photon sources are yet to be interfaced with an absorptive quantum memory.  

\item {\bf Atomic heralded photon sources:} Emissive quantum memories, as proposed by Duan, Lukin, Cirac and Zoller~\cite{Duan2001}, generate correlated pairs of Stokes and anti-Stokes photons~\cite{Nature.423.731,Science.301.196}. One of the generated photons can then be stored in a second (absorptive) quantum memory. Because the source of the single photons (the emissive memory) is perfectly matched to the target (absorptive) memory, the photons are well-matched for storage~\cite{Chaneliere2005,nature.452.67,Eisaman2005}. While these techniques have been pioneered using clouds of ultracold rubidium, the principle can be applied to many other systems. By harnessing slow-light effects in double-$\Lambda$ EIT systems the temporal duration of the generated photon pairs has been tuned between 50 and 900\,ns~\cite{Balic2005,Du2008}, however such photons have not yet been integrated with a quantum memory. 

\item {\bf Broadband Raman memory:} Far-off-resonant Raman memories have been shown to store THz-bandwidth light~\cite{PhysRevLett.111.083901,PhysRevLett.111.243601}; this bandwidth is directly compatible with SPDC photon sources. The diamond Raman memory has been used to store single photons with a 1.7\,THz bandwidth~\cite{PhysRevLett.114.053602}.
\end{enumerate}


\subsubsection{Telecom wavelength storage}
Global communication networks are based on silica fibers in which the attenuation is minimum between $\sim1.2$\,$\mu$m and $\sim1.6$\,$\mu$m. This window is split by a strong OH$^-$ absorption peak at $\sim1.4$\,$\mu$m leading to three popular wavelength ranges for telecommunications: The \emph{O-Band} (1260 to 1360\,nm), The \emph{S-Band} (1460 to 1530\,nm), and the \emph{C-Band} (1530 to 1560\,nm). The S and C bands are the most widely used due to the availability of Erbium-doped fiber amplifiers, but the O-Band is also popular as it has close to zero dispersion. Future long-range quantum communication systems will have to run alongside conventional communications, using the same infrastructure. Therefore, there is a clear motivation to develop quantum memories operating in the telecommunication wavelength bands for use in quantum repeaters. However, this is not a necessary requirement as we discuss in~\ref{alt_telecom}. It should be noted that many of the alternative applications for quantum memories discussed in this review do not involve long distance communication, and therefore do not need to operate at telecommunications wavelengths. For these applications other wavelength considerations, such as the efficiency of detectors or the availability of suitable lasers, may be of greater importance.

Far-off resonant Raman-type memories~\cite{PhysRevLett.111.083901,PhysRevLett.111.243601,PhysRevLett.114.053602} are broadly tunable and therefore could, in principle, operate at a vast range of wavelengths, including in the telecom band. However, at the time of writing, the storage times achieved in these systems are not suitable for communications. Most quantum memories with lifetimes compatible with communications have utilized near-resonant optical transitions in atomic vapors or cryogenic solids. The challenge for telecom wavelength storage is therefore to find a suitable memory substrate, with a long-lived metastable state, in which a dipole-allowed transition from the ground state exists in the telecom band. To date, the only system in which telecom band storage has been demonstrated is the erbium ion (Er$^{3+}$) which absorbs light within the C-Band~\cite{NatPhoton.9.83}. However, inefficient optical pumping and associated noise processes introduce difficulties for quantum level storage. Motivated by the difficulties in direct storage of telecom light, a number of indirect methods have been demonstrated, including frequency conversion and teleportation, to interface quantum memories with telecom-wavelength single photons. A summary of the direct, and indirect, methods for interfacing quantum memories with telecom-wavelength photons is given in the sections below followed discussion of an alternative approach that bypasses the need for telecom wavelength storage.

\paragraph{Direct storage of telecom wavelengths in Er$^{3+}$}

The first proof-of-principle demonstration of light storage in Er$^{3+}$ used the controlled reversible inhomogeneous broadening (CRIB) protocol in an Er$^{3+}$:Y$_2$SiO$_2$ crystal cooled to 2.6\,K~\cite{Lauritzen2010,Lauritzen2011}. A switchable external electric field gradient was used to apply reversible inhomogeneous broadening via the linear Stark effect. Weak pulses of $\approx200$\,ns duration and mean photon number $\bar{n} = 0.6$ were stored for up to 600\,ns, with a peak efficiency of 0.25\%. The efficiency and lifetime  of optical storage in Er$^{3+}$:Y$_2$SiO$_2$ was dramatically improved using the revival of silenced echo (ROSE) scheme to store pulses for 16\,$\mu$s with 42\% efficiency~\cite{Dajczgewand2014}. However, this scheme is yet to be tested at the single photon level where performance may be limited by spontaneous emission noise. The AFC protocol has also been applied to Er$^{3+}$ where a cryogenically-cooled erbium-doped fiber was used to store heralded single photons for a fixed delay of 5\,ns with efficiency 1\%. The photons are generated in correlated pairs by spontaneous parametric downconversion with the detection of a photon at 795\,nm used to herald the presence of a signal photon at 1532\,nm, which is stored in the memory. Both time-bin entanglement between the herald and signal photons~\cite{NatPhoton.9.83} and polarization state of the signal photon~\cite{PhysRevLett.115.140501} have been preserved during storage in the memory.

\paragraph{Indirect storage of telecom wavelengths}

While no accessible optical transitions in the telecom bands exist from the ground state of the alkali metals, the $\ket{5P_{3/2} F=2} \longrightarrow \ket{6S_{1/2} F=1}$ transition in rubidium occurs at a convenient wavelength of 1367\,nm. This transition was employed by Radnaev {\em et al.} to convert the frequency of a single photon from 780\,nm to 1367\,nm via resonantly-enhanced four-wave mixing~\cite{NatPhys.6.894}. The single photons at 780\,nm were generated by anti-Stokes scattering of a write beam in an emissive quantum memory in an optical lattice containing ultracold $^{87}$Rb. Frequency conversion to 1367\,nm was achieved in a second ultracold $^{87}$Rb cloud with efficiency 55\%, before the light was converted back to 780\,nm by the reverse process. Finally, the coherence in the emissive memory is read out via Stokes scattering of a read beam to produce a photon at 795\,nm. Memory lifetime of up to 0.1\,s was achieved and the non-classical photon statistics of the anti-Stokes beam were preserved during frequency down- and up-conversion. Entanglement between the frequency-converted qubit and the emissive memory was also reported~\cite{PhysRevLett.105.260502}. 

The use of an atomic transition restricts the wavelength and bandwidth of the frequency conversion; greater flexibility can be achieved by using non-linear optics. While non-linear optics has traditionally required extremely intense pulsed lasers~\cite{Boyd2003}, the development of periodic poling~\cite{Houe1995} and waveguides in non-linear optical materials~\cite{Poel1990} has opened the door for non-linear optics using single photons~\cite{Tanzilli2005,PhysRevLett.105.093604,Rakher2010,Rakher2011,McKinstrie2012,NatPhoton.7.363,Guerreiro2014}. These techniques have been used to interface quantum memories with telecoms fiber. Albrect {\em et al.} used a periodically-poled lithium niobate (PPLN) waveguide to convert heralded single photons from an emissive rubidium memory to 1552\,nm via difference frequency generation with efficiency 14\%~\cite{Albrecht2014}.The opposite approach was taken by Maring {\em et al.} who up-converted telecom photons at 1570\,nm to 606\,nm via sum frequency generation in a periodically-poled potassium titanyl phosphate (PPKTP) waveguide with efficiency 22\%. The up-converted photons were then stored in a Pr$^{3+}$:Y$_2$SiO$_2$ quantum memory~\cite{Maring2014}.

\paragraph{Alternatives to telecom storage}\label{alt_telecom}

A clear motivation for the storage of telecom-wavelength single photons is for long-range entanglement distribution~\cite{Simon2007}. The role of the memory in these schemes is to act as a temporal buffer to overcome the inherently probabilistic nature of the photon pair generation. To maximize the distance between stations, it is desirable that the photons be in the telecom band; hence the motivation for quantum memories operation at these wavelengths. However, it is worth noting that only one photon from each pair source needs to travel a long distance. Therefore, if an entangled pair source is constructed where one photon is in the telecom band and the other is, for example, in the visible, then the telecom photon can be coupled into optical fiber while the visible one is locally stored in the quantum memory. This removes the requirement for quantum memories to operate at telecom wavelengths, at the expense of increasing the number of repeater stations required per unit length. A summary of developments in this area is included in the following paragraph.

A cavity-enhanced SPDC photon pair source was developed by Fekete {\em et al.} in which one of the photons has a wavelength of 1436\,nm and the other 606\,nm, which is compatible with a Pr$^{3+}$:Y$_2$SiO$_5$ AFC quantum memory~\cite{Fekete2013}. Critically the cavity produced photons of extremely narrow linewidth ($\sim2$\,MHz) which meets the stringent requirements of the memory. The 606\,nm photons were subsequently stored in the quantum memory with on-demand readout up to 4.5\,$\mu$s~\cite{Rielander2014}. A similar experiment was performed by Clausen {\em et al.} where one of the down-converted photons was of wavelength 1338\,nm and the other, at 883\,nm, was suitable for storage in a Nd$^{3+}$:Y$_2$SiO$_2$ AFC memory. The 120\,MHz bandwidth required for the memory was achieved by strong spectral filtering of the signal and herald photons. The 883\,nm photon was stored for fixed delays of up to 200\,ns and time-bin entanglement between the memory output and the telecom wavelength photon was demonstrated~\cite{nature.469.508}. Subsequently, polarization qubit storage was also demonstrated in this system by using a pair of orthogonally-aligned cyrstals~\cite{PhysRevLett.108.190503}. In an extension to this work, Bussi\'eres {\em et al.} achieved the teleportation of the quantum state of a flying qubit into the quantum memory~\cite{Bussieres2014} despite over 25\,km of optical fiber separating the memory from the source of flying qubits.

\subsubsection{Integrability of implementations (Nano/micro fabricated systems)}
 \label{sec:Integration}
The development of quantum memories in waveguides or optical fibers is motivated by the desire to integrate memories into quantum photonic circuits~\cite{OBrien2007,Science.339.798} or fiber-optical quantum communication networks~\cite{Hubel2007}. Furthermore, by providing tight modal confinement over long interaction lengths, optical waveguides/fibers can offer a significant increase in memory efficiency over bulk systems and reduce the power requirements for the read and write fields. Of all the quantum memory substrates, rare-earth ion-doped crystals and glasses are the best suited for integration as waveguides and optical fibers are widely available. Indeed many rare-earth ion schemes have taken advantage of this feature (see refs.~~\cite{PhysRevLett.98.113601,NatPhoton.9.83,Jin2015,Saglamyurek2012,nature.469.512,PhysRevLett.113.053603}). 

The development of hollow-core photonic crystal fibers (HCPCFs), which can be filled with molecular gases~\cite{benabid:123903,science_298_399,Ghosh2005} or atomic vapours~\cite{Perrella2013,Londero2009}, has opened the door to a plethora of non-linear optics experiments using low-intensity light~\cite{Bhagwat2008}. These results are promising for integrated quantum memories using atoms and molecules in hollow-core fibers; however, significant roadblocks still have to be overcome. Alkali metal vapors are the most widely used substrate in bulk quantum memories, and hence are the logical choice for filling hollow core fibers; unfortunately, alkali metal atoms adsorb onto glass surfaces so high optical depth cannot be achieved, even timescales of a year. This problem was circumvented by Slepkov and co-workers~\cite{Slepkov2008} who showed that an intense pulse of light could be used to blast atoms off the internal surface of the fiber. This technique, know as light-induced atomic desorption (LIAD), can transiently increase the optical density in the fiber by orders of magnitude. Using the LIAD technique, Sprague {\em et al.} loaded Cs vapor into a Kagome-structured HCPCF~\cite{Benabid2011} with diameter 26\,$\mu$m; they achieved optical depths in excess of 3000 and demonstrated the optical preparation reqired for storage~\cite{NJP.15.055013}. Following the cesium Raman memory protocol~\cite{Reim2010} they then stored GHz-bandwidth pulses with an aveage of 1.8 photons per pulse and a signal-to-noise ratio of 2.8:1 at the single-photon level. The read/write pulse energy required for the fiber-based memory was 200 times lower than that used in the vapor cell~\cite{Reim2010}, illustrating the advantage of the integrated platform. The memory operation was hampered by highly-transient nature of the optical density which decays on a 30\,s timescale after the LIAD beam is switched on and requires several hours without LIAD to recover peak optical depth. However, in a follow-up work, Kaczmarek {\em et.al.} were able to achieve persistently high optical depths for over 80 days with promising implications for future memory experiments~\cite{Kaczmarek2015}. 

Despite the obvious advantages of the HCPCF for atomic quantum memories, there is also a significant disadvantage: collisions with the wall of the fiber can cause a spin-flip in the atom resulting in a loss of coherence across the ensemble. Therefore, the \textsl{transit time} of an atom/molecule across the diameter of the fiber is a limiting factor in the lifetime of the memory. At room temperature, this limits the lifetime of the cesium HCPCF memory to $\sim100$\,ns compared to over 1\,$\mu$s in bulk. The transit time can be increased by loading ultracold atoms into the fiber~\cite{Christensen2008} and has been used to achieve slow-light effects~\cite{Bajcsy2009}; however, to date, a quantum memory has not been demonstrated in such a system. Instead, efforts to deliver ultracold quantum memories on an integrated platform have focused on using tapered nanofibers~\cite{Tong2003} --- sub-wavelength diameter waveguides in which a significant fraction of the optical mode propagates outside of glass. Nanofibers are fabricated by heating and drawing standard optical fiber and hence are easily coupled to conventional fiber optics. Light storage using a 400\,nm diameter tapered fiber surrounded by a cesium magneto-optical trap (MOT) was demonstrated by Gouraud {\em et al.}~\cite{PhysRevLett.114.180503}. An optical density of 3 was achieved from an estimated 2000$\pm$500 atoms in the interaction region. Single-photon-level pulses ($\bar{n} = 0.6$) were stored with 10\% efficiency at a signal-to-noise ratio of 20:1. The temperature of the MOT was 200\,$\mu$K resulting in a transit time of 3.5\,$\mu$s which is the dominant factor in determining the memory lifetime. Similar work was performed by Sayrin {\em et al.}~\cite{Optica.2.353} who trapped ultracold cesium atoms in an optical lattice surrounding a tapered nanofiber. The optical lattice forms a series of nano-traps 225\,nm from the fiber such that the atoms are closer to the fiber than in the MOT. This resulted in a remarkable optical depth of 5.9 from just 160 atoms. An EIT scheme was used to slow light down to 50\,m/s and to store pulses containing $\bar{n} = 0.8$ photons with 3\% efficiency for 2\,$\mu$s. In closely-related work, the authors have demonstrated that millisecond coherence times can be achieved in the magnetic-field-insensitive ground states of cesium in the dipole trap~\cite{PhysRevLett.110.243603}. These exciting results suggest that much higher memory lifetimes and efficiencies may be feasible using nanofibres in atomic clouds, but this has yet to be demonstrated.

\section{Emerging applications for quantum memories}\label{sec:application}


\subsection{Optical signal processing}\label{sec:Processing} 


Light-matter interfaces are one of the key elements in the implementation of photonic quantum information processing. Effective light-matter interactions require careful engineering of matter and optical modes to enable high-fidelity operation. This imposes limitations on the central wavelengths, bandwidths, and temporal profiles of the optical modes. More sophisticated architectures will therefore benefit from optical signal processors as interconnects for controllably manipulating these modes. Such manipulations of quantum optical signals must be noiseless and coherent in order to preserve the quantum properties of the optical state. 

The field of quantum frequency conversion~\cite{Kumar1990a,Raymer2012} has advanced rapidly in recent years with techniques such as four-wave-mixing~\cite{PhysRevLett.105.093604}, sum-frequency~\cite{NatPhoton.7.363,Guerreiro2014,Rakher2010,Rakher2011,Tanzilli2005,Langrock2005} and difference-frequency generation~\cite{Pelc2012,NatureComm.2.537} affording control over the bandwidth and carrier frequency of single photons. However, full control over the temporal modes of a quantum optical state requires buffering; {\em i.e.} a quantum memory. In applications that require both temporal and frequency mode manipulation ({\em e.g.} refs. \cite{PhysRevLett.113.053603}, \cite{Humphreys2014} or \cite{Campbell2014}) it is interesting to explore the option of using a quantum memory to store and manipulate different frequency modes offering temporal and spectral signal processing on a single platform. In the following paragraphs, the role of quantum memories in frequency and temporal mode manipulation are outlined.


In ~\cite{nature.461.241}, the authors used a Raman-coupled gradient echo memory in a Rb vapor cell to demonstrate a coherent optical pulse sequencer. They stored a sequence of input optical signals in an ensemble of Rb atoms, in which the energy of the atomic level used for storage was longitudinally broadened along the vapor cell by a spatially varying magnetic field. This field induced inhomogeneous broadening leads to dephasing of the collective atomic excitations. However, this can be reversed by inverting direction of the magnetic field gradient, resulting in rephasing of quantum state of the atomic ensemble. Each time the magnetic field is switched, a rephasing is expected at a later time. If the number of magnetic field switches between storage and retrieval is even, stored input signals get rephased with the same time ordering in which they arrived at the input of the medium. Applying a control field at this point then allows for the retrieval of the input signals. This is called first-in-first-out (FIFO). However, retrieving the input sequence after an odd number of magnetic field switches results in first-in-last-out (FILO), as the first (last) input gets rephased later (earlier) than the rest of the sequence. Combining these two techniques allows one to arbitrarily re-order optical signals \cite{nature.461.241}. Although this experiment was performed on weak coherent pulses, the same technique can be applied to process single photon pulses.

Careful optical mode-matching is essential for quantum information transfer between systems. Building the ability to coherently manipulate optical modes into optical quantum memories would remove the need for mode-manipulation elements and the complexity and loss associated with them. In \cite{PhysRevX.2.021011}, a Raman-coupled gradient echo memory was employed for precise spectral manipulation of optical pulses. Controlled inhomogeneous broadening due to the external magnetic field was used not only for storage and retrieval but also for central frequency manipulation, bandwidth manipulation, spectral filtering and pulse interference. This was performed by precisely controlling the magnetic field strength along the medium with eight separate solenoid coils. Given that the inhomogeneously broadened spectrum for storage and manipulation is associated with the length of the medium, a larger bandwidth or an attempt to accommodate more pulses leads to a reduction in the effective optical depth and therefore in efficiency. This can be compensated through spatial multiplexing for more practical implementations. In addition, more sophisticated control and functionalities may become possible with spectral or temporal modulation of the Raman control field.

In a more recent experiment \cite{NJP.16.065019}, the authors employed a rare-earth-ion-doped crystal and an AFC storage protocol for processing optical signals. The AFC was implemented in a Ti-diffused Tm:LiNbO$_3$ waveguide, which has the potential for more compact and integrable memories. The medium provides a broad absorption spectrum that allows one to simultaneously prepare multiple AFC channels with different comb spacings (rephasing times). Using a phase modulator, one can then map each incoming signal into a specific AFC. This enables pulse sequencing, along with time-to-frequency multiplexing and demultiplexing. In addition, the authors demonstrated pulse compression and stretching. This relies on linearly chirping the input pulse using a phase modulator, along with storage in an AFC with varying peak spacing. The linear chirp rate and the gradient in the peak spacing ultimately determine the output pulse duration. All of these features promise a versatile and yet integrable quantum optical signal processor for photonic quantum information processing architectures.

Raman coupling at light-matter interfaces offers an alternative route to spectral and temporal selectivity. Temporal engineering of the Raman coupling underlies the Raman quantum memory protocol \cite{Nunn2007,PhysRevA.76.033805}, which is the basis for the room-temperature implementation of broadband quantum memories \cite{Reim2010a,Reim2011PRL,PhysRevLett.111.083901,PhysRevLett.114.053602}. Temporal mode selectivity in Raman based systems led to the proposal and demonstration of multipulse addressing of a Raman quantum memory for optical state engineering and efficient readout \cite{PhysRevLett.108.263602}. This was shown using a Raman memory in warm cesium vapour. The multipulse readout results in a network of configurable beam splitters acting on a set of temporal modes. Such a system can be used to prepare high-dimensional time-bin single-photon states for quantum key distribution \cite{Nature.509.475}.

Depending on the Raman medium, a large detuning from single photon transitions can be used to provide a broad spectral range for wavelength selectivity while two-photon resonance is maintained by tuning the control field. This feature has been used in~\cite{Fisher2015} to store heralded single-photon pulses at 720~nm and shift the wavelength of the recalled pulse by approximately $\pm$9\,nm (18~nm range). The authors also demonstrate bandwidth manipulation via pulse stretching and compression over a 3.75\,THz range while preserving non-classical correlations.


\subsection{Optical quantum computation}\label{sec:Computing}

\subsubsection{Linear optical quantum computation}

Several different physical implementations for quantum information processing are under active research. Of these, quantum memories have a critical role to play in the context of optical quantum information processing, which focuses on manipulating quantum information encoded in degrees of freedom of the electromagnetic field. In particular, many approaches to optical quantum information processing focus on the manipulation of single-photon-encoded qubits~\cite{ralph2010optical,RevModPhys.79.135}. The central challenge of these schemes is that two-qubit gates require the ability to induce a strong non-linearity between optical modes containing only a few photons, which is experimentally challenging~\cite{chang2014quantum}. Fortunately, Knill, Laflamme and Milburn showed that it was, in principle, possible to carry out quantum information processing using only linear optical elements (beam splitters and phase shifters), single photon sources and photo-detectors~\cite{Knill2001}. In this case, two-qubit gates are induced probabilistically by employing the effective non-linearity induced by single photon sensitive detectors. This shifts the experimental burden from the implementation of strong optical non-linearities to the generation of a large number of ancillary resources, necessary to overcome the probabilisitic nature of the two-qubit gates~\cite{ralph2010optical,Li2015b}. Furthermore, as is the case for other optical schemes, the initial generation of single photon states typically also relies on probabilisitic processes. These drawbacks mean that scaleable implementations of LOQC beyond a few qubits have proven challenging.

There are two aspects in which quantum memories can have a significant impact on the prospects for implementing scalable LOQC. First, quantum memories allow for the synchronisation of multiple independent non-deterministic operations. In essence, this is achieved by allowing each operation to be repeated until it is successful, and then storing the outputs of the operation in quantum memories until they are needed. This synchronisation is vital for the implementation of more complex sequences of quantum gates \cite{RevModPhys.79.135}. In addition, as shown in \cite{PhysRevLett.110.133601}, by applying this technique to single-photon sources, quantum memories can significantly enhance the rate of generation of synchronized single photons from arrays of heralded probabilistic single-photon sources~\cite{PhysRevLett.110.133601}. 

\begin{figure} 
\center{\includegraphics[width=1.0\linewidth]{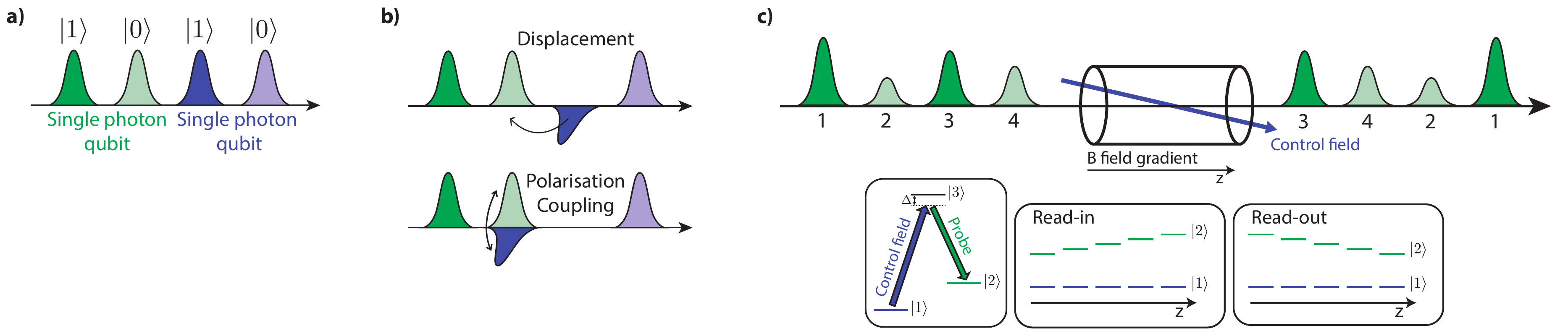}}
\caption{a) Two single photon qubits are encoded in successive time bins within a single spatial mode. b) LOQC can be carried out by controllably reordering and coupling different time bin modes. Without only controllable birefringent elements, this requires time-bins to be rotated into the orthogonal polarisation to be displaced, and only one displacement can be implemented at each step. c) A pulse sequencer, as can be achieved using a gradient echo memory, allows for arbitrary re-ordering of time bins.}
\label{fig:TimeBinConcept}
\end{figure}

Second, as has been discussed in the previous section, quantum memories are able to selectively interface with specific time-frequency modes of the electromagnetic field. Moving beyond their use in simply synchronising optical modes, this unique ability allows them to be employed as versatile quantum information processing elements. This utility becomes clear when considering time-encoded optical quantum information processing. Historically, linear optical quantum computing schemes have primarily been described in terms of spatial or polarisation encodings, in which two spatial or polarisation modes are required for each qubit. In \cite{HumphreysPRL2013}, the authors instead considered using a string of time-bin encoded qubits within a single spatial mode and a set of time-dependent linear optical operations to implement LOQC by reordering and coupling time-bins (Fig.~\ref{fig:TimeBinConcept}a\&b). Such time-frequency encodings are advantageous due to the large space of available modes within even a single spatial mode, which may enable information processing protocols to be simplified and made more compact. The number of individual operations in the proposed scheme can be significantly reduced if a processing element can allow for the arbitrary re-ordering of several time-bins. This can be performed by using an optical pulse sequencer, which has been demonstrated for classical pulses in a warm-vapor gradient echo memory (Fig.~\ref{fig:TimeBinConcept}c)~\cite{nature.461.241} and in a multi-mode quantum memory based on a rare-earth-ion doped crystal \cite{NJP.16.065019}. 

An additional degree of freedom unique to some quantum memories is their frequency selectivity. This allows for interactions with optical modes in which both the temporal profile and the central frequency of the coupled mode can be dynamically controlled \cite{Campbell2014,Humphreys2014}. In \cite{Campbell2014}, the authors exploit this frequency selectivity to propose a method to use quantum memories for implementing configurable unitary transformations and linear logic gates on frequency-encoded quantum information. This is achieved using a set of off-resonant Raman-coupled quantum memories; by changing the frequency of the strong control field, the coupled optical mode can be accordingly altered. The authors demonstrate how their scheme could be employed to to implement a conditional controlled-z gate. They further show that the scaling of the number of quantum memories is favorable, where for an arbitrary $N$-mode unitary operation, $N$ quantum memories are sufficient. 

The ability of quantum memories to act as versatile elements for the manipulation of time-frequency modes may also find an application in demonstrations of boson sampling. This restricted form of quantum information processing utilises only single photons and linear optical manipulations~\cite{Science.339.794,Science.339.798,tillmann2013experimental}, and yet, surprisingly, is able to carry out a sampling problem that is strongly believed to be exponentially hard for classical computers~\cite{Aaronson}. Several recent papers have considered using time- and frequency-selective elements for boson sampling~\cite{Pant2015,Motes2014a}. These protocols could be more compactly implemented using quantum memories based on the techniques highlighted in this section.

\subsubsection{Photonic cluster states for one-way quantum computing}
Measurement-based (one-way) quantum computing (MBQC) is an efficient form of LOQC that relies on a pre-prepared many-body entangled resource state, a \textsl{cluster state}~\cite{BrownePRL2005}. Once this resource state has been prepared, single-qubit measurements are sufficient to perform quantum computing tasks \cite{Raussendorf2003}. Optical implementations of MBQC therefore require the scalable preparation of photonic cluster states. In \cite{BrownePRL2005}, the authors introduced a set of linear optical operations for the probabilistic preparation of such states. Experimental preparation of photonic cluster states and demonstrations of one-way quantum computing have been achieved for a limited number of qubits \cite{Lu2007,walther2005experimental}. As is the case with other linear optical quantum information protocols, scalable photonic implementations of MBQC would also benefit from using quantum memories to enhance the success probability of the cluster state generation \cite{BodiyaPRL2006}.

Thus far, we have focused on single-photon-encoded quantum information. As an alternative, information can be encoded in the quadrature amplitudes of an optical mode~\cite{Braunstein2005}. Unlike the finite-dimensional information that can be encoded in the modal occupation of single photons, the quadratures present a continuum of eigenstates in which information can be encoded. This allows for a different model of information processing, termed continuous variable (CV) quantum information processing \cite{andersen2010continuous}. In \cite{MenicucciPRL2006}, the authors generalize the notion of discrete-variable one-way quantum computation to continuous-variable cluster states. They propose an optical implementation using squeezed-light sources, linear optics, homodyne detection and single-mode non-Gaussian measurement; see also \cite{ZhangPRA2006} and \cite{MenicucciPRA2007}. Further progress has been made with preliminary experimental demonstrations \cite{SuPRL2007,YukawaPRA2008} and more advanced theoretical proposals \cite{menicucci2008one,menicucci2010arbitrarily,menicucci2011temporal}, which led to ultra-large-scale generation of time-multiplexed continuous-variable cluster states \cite{yokoyama2013ultra}.

\begin{figure} 
\center{\includegraphics[width=0.9\linewidth]{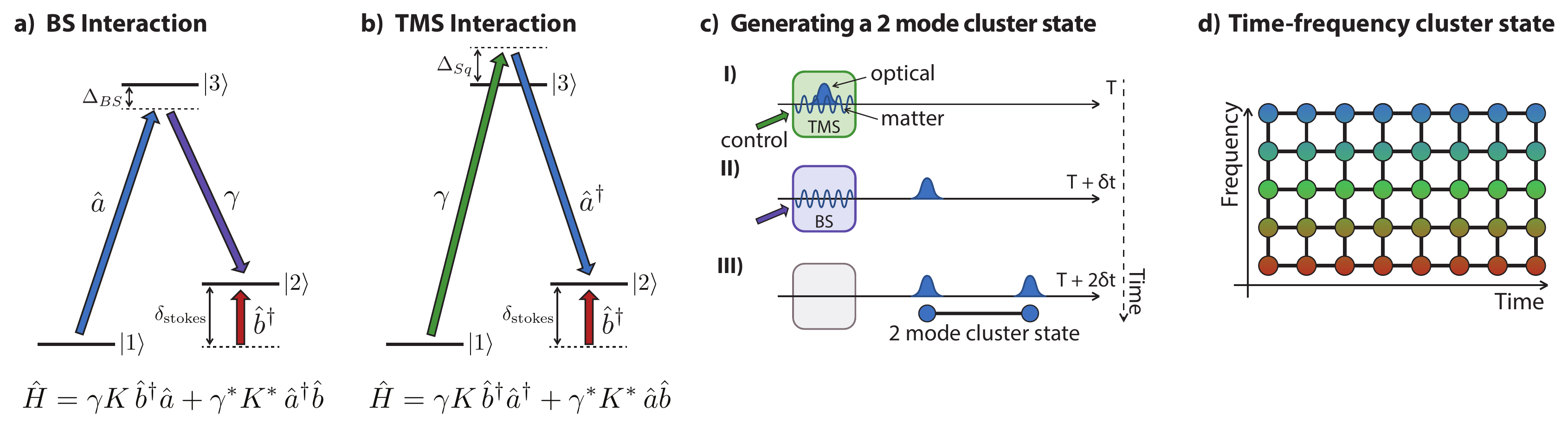}}
\caption{An example off-resonant Raman memory scheme based on a three-level system. An electromagnetic field mode $\hat{a}$ is coupled to the memory transition $\hat{b}$ between levels $\ket{1}$ and $\ket{2}$ through a Raman transition to a virtual energy level detuned by $\Delta$ from $\ket{3}$. By adjusting the frequency of the control field-mode $\gamma$, both a) beam splitter interactions and b) two-mode squeezing interactions can be generated between the modes. The BS operation can realise read in/out from the memory. c) Using these interactions, entangled 2 mode cluster state primitives can be generated. d) These clusters can be stitched together using further quantum memories to create a time-frequency continuous-variable cluster state encoded in modes tiled across the available time and frequency degrees of freedom.}
\label{fig:LambdaSystem}
\end{figure}

In an attempt to improve the scalability of optical CV quantum computing, the authors in \cite{Humphreys2014} propose the use of quantum memories for quantum computing in optical time-frequency multiplexed CV cluster states. In this scheme, the previously separately discussed time- and frequency-selectivity of Raman quantum memories are combined in order to address states within a two-dimensional space of time-frequency modes (Fig~\ref{fig:LambdaSystem}). The proposal further takes advantage of two types of operations that are available through Raman interactions in order to generate and manipulate time-frequency cluster states. The first operation is a beam splitter interaction between an optical mode and the memory mode (Figure~\ref{fig:LambdaSystem}a), where the interaction Hamiltonian is given by ${\hat H}=\gamma\kappa{\hat b}^{\dagger}{\hat a}+\gamma^*\kappa^* {\hat a}^{\dagger}{\hat b}$. This is the standard Hamiltonian that underlies Raman quantum memories. The second less commonly employed operation is a two-mode squeezing interaction (Figure~\ref{fig:LambdaSystem}b), which can be described by ${\hat H}=\gamma\kappa{\hat b}^{\dagger}{\hat a}^{\dagger}+\gamma^*\kappa^* {\hat a}{\hat b}$. This interaction allows entanglement to be generated between the memory mode and an optical mode. Using a combination of these two interactions, cluster states encoded in $d$ frequency modes and an arbitrary number of time-bin modes can be generated using only $7d-3$ quantum memories (Figure~\ref{fig:LambdaSystem}c\&d).

\subsection{Optical non-linearity mediated by storage}\label{sec:nonlinear}
Creating non-linear interactions at the single photon level has been a great challenge in quantum optics that requires unprecedented tailoring of light-matter interfaces~\cite{chang2014quantum}, where quantum storage techniques may become useful for mediating photon-photon interaction. Successful implementations of optical non-linearity enable photonic two-qubit gates \cite{NemotoPRL2004} and non-destructive Bell-state detection \cite{BarrettPRA2005}, and photon switching \cite{PhysRevLett.81.3611, PhysRevLett.108.227402, PhysRevLett.113.053602, chen2013all, tiecke2014nanophotonic}. 

In the seminal work of Schmidt and Imamo\u{g}lu \cite{OptLett.21.1936}, cross-phase modulation based on the AC Stark effect and phase sensitivity of EIT have been employed to a create large Kerr non-linearity. In several experiments this scheme and its variations were used to induce non-linear interactions between weak propagating electromagnetic fields \cite{chen2006low,LiPRL2008,FeizpourXPM2015,PhysRevLett.116.033602}. The causal, non-instantaneous nature of the cross-Kerr interaction between propagating photonic pulses has been shown to preclude its application in photonic quantum information processing~\cite{shapiro2006single,shapiro2007continuous}. Quantum storage of at least one photonic mode can circumvent this limitation~\cite{Parades-BaratoPRL2014,beck2015large,tiarks2015optical}.

In \cite{Hosseini2012}, the authors employed a quantum memory to enhance cross-phase modulation between signal and probe fields. In their proposal both fields are simultaneously stored using a Raman-coupled gradient echo memory \cite{ncomms.2.174} in a $\Lambda$ system, where the AC Stark shift due to the signal results in cross-phase modulation on the probe field. In a proof-of-principle experiment with a freely propagating signal and a stored probe, they demonstrated a cross-phase modulation of about $10^{-12}$ rad per signal photon \cite{Hosseini2012}. An impressive conditional cross-phase shift of up to $\pi/3$ has been demonstrated in a very recent development by Beck {\it et.al.}~\cite{beck2015large}. In their experiment, a cold ensemble of $^{133}$Cs atoms inside a high finesse cavity were used to store a signal photon using EIT, where the cavity significantly enhances the cross-phase shift due to a control photon in resonance with the cavity (with average photon number of $\langle n_c\rangle\leq 0.5$)~\cite{beck2015large}.

Achieving single photon sensitivity requires strong lateral confinement. This can be pursued via light-matter interfaces based on waveguides in solids \cite{nature.469.512,PhysRevLett.115.013601}, and gas-filled hollow-core photonic crystal fibers \cite{NJP.15.055013,venkataraman2013phase}. In a recent experiment, Sinclair {\it et al.} demonstrated cross-phase modulation on a probe stored in a thulium-doped lithium niobate waveguide, using the atomic frequency comb technique~\cite{sinclair2015cross}. In their scheme, a spectral pit (transparency window) is prepared for the signal to propagate through the medium and have off-resonant interaction with the stored probe. In this proof-of-principle experiment a cross-phase shift of about 10$^-9$ per signal photon has been shown. Storage of the probe allowed the authors to demonstrate cross-phase shifts that are insensitive to arrival time of the signal field, and therefore is a step towards quantum non-demolition detection of photonic time-bin qubits. 

Reiserer {\it et al.} achieved single-photon non-linearity using a different scheme that does not rely on off-resonant interaction and is thus immune to absorption loss as opposed to cross-phase modulation based on the AC Stark effect~\cite{reiserer2013nondestructive}. They demonstrated non-destructive detection of single photons using single atoms trapped in a cavity. The same system was later used to apply a quantum gate between a flying optical photon and a single trapped atom~\cite{reiserer2014quantum}. This implementation, comprising a single atom trapped in a cavity, was successfully used to demonstrate quantum storage of optical photons~\cite{nature.473.190} and heralded storage of photonics qubits~\cite{kalbPRL2015}. Despite the exceptional properties of atom-cavity light-matter interfaces, for practical reasons, it is advantageous to pursue other quantum memory implementations for mediating or implementing optical non-linearity. 

Strongly coupled quantum dot-cavity systems have also been shown to be very effective in observing optical non-linearity between weak pulsed and CW fields. Several groups demonstrated quantum dots strongly coupled to micro/nano-cavities~\cite{PhysRevLett.108.227402,PhysRevLett.108.093604,PhysRevLett.109.166806}. Practical implications of fast operation rates and potential for on-chip integration of these devices are appealing with potential applications in quantum information processing.

Another promising avenue for creating optical non-linearity is based on strong dipole-dipole interactions between high-lying Rydberg states of alkali atoms~\cite{SaffmanRMP2010, dudin2012strongly, peyronel2012quantum}. This led to a great deal of experimental and theoretical research, to use a Rydberg blockade or Rydberg-induced phase shift for applications in photonic quantum information processing. Recent proposals~\cite{Parades-BaratoPRL2014, KhazaliPRA2015,das2015photonic} use storage based on EIT to decouple light propagation and Rydberg interaction for practical implementations of Rydberg-based photon-photon gates. In the most recent and state-of-the-art experiment, Tiarks {\it et.al.}~\cite{tiarks2015optical}, demonstrated optical $\pi$ phase shift created by a single-photon pulse. In their experiment, a control single-photon pulse with an average photon number of 0.6 is stored in a high-lying Rydberg state in a cold ensemble of $^{87}$Rb atoms. Then a target photon pulse with average photon number of 0.9 propagates via Rydberg-EIT, where the target photon experiences a controlled phase shift of up to $\pi$. This is a significant step towards the long-standing goal of deterministic photon-photon gate.

In \cite{RispePRL2011, VoPRL2012}, an alternative physical process has been considered to mediate photon-photon interaction that relies on atomic collisions in a Bose-Einstein Condensate (BEC) to realize a controlled-phase gate between two stored photons. Despite the fact that the collisional interactions are very weak, these approaches take advantage of the long storage time of about 1 second in BECs~\cite{PhysRevLett.103.233602}. Such a Kerr non-linearity has also been used to study the realization of macroscopic coherent spin states, where stored coherent states evolve into so-called cat states due to atomic collisions in BEC~\cite{LauPRL2014, TrailPRA2014}.

\section{Summary and outlook}
Photonic quantum memory research has seen a wealth of progress on multiple fronts in recent years, as the considerable citation list will attest. Memories can be conceived as controllable light-matter interfaces for building elements in optical quantum information processing. Their applications now extend~\cite{Bussieres2013} far beyond their initial role in long-distance quantum communication based on linear optics~\cite{RevModPhys.83.33}.  In this review, we focused on recent experimental and theoretical developments in implementing photonic quantum memories and their emerging applications. 


The demands placed on a quantum memory will vary widely, depending on the desired application. Many experimental implementations offer excellent performance in several key areas; however, technical challenges remain to be overcome for practical use of most examples, whether in storage lifetime; signal-to-noise ratio; duty cycle; efficiency; or surfeit complexity, for example. The impressive progress of recent years gives a strong reason to believe that these challenges will be surmounted. 

In addition, it is desirable to design implementations that allow wavelength tunability, and the possibility of spectral- and spatial-multiplexing. Incorporating multiple functionalities, such as precision temporal and spectral control, or wavelength conversion, into memory design will enable more versatile elements for photonic quantum information processing architectures. This is particularly important given the wavelength- and bandwidth-specificity of other quantum components such as sources and detectors. Moreover, implementations with robustness, ease-of-use, and potential for integration are likely to find favour for practical applications.

\section*{Acknowledgements}
We thank Erhan Saglamyurek for fruitful discussions and Ilja Gerhardt, Rados{\l}aw Chrapkiewicz, Micha{\l} Dabrowski, Sergey Moiseev, and Daniel Higginbottom for their helpful comments. PH was supported by the UK EPSRC Hub for Networked Quantum Information Technologies (NQIT), JN acknowledges a Royal Society University Research Fellowship, and BS acknowledges support through funding from NSERC.

\bibliographystyle{unsrt}

\end{document}